\documentclass[]{spie}  %>>> use for US letter paper
\usepackage{amsmath,amsfonts,amssymb}
\usepackage{graphicx}
\usepackage{xcolor}
\usepackage{booktabs}
\definecolor{tealblue}{rgb}{0.18, 0.40, 0.46}
\definecolor{ruby}{rgb}{0.88, 0.07, 0.37}
\usepackage[pagebackref=true,breaklinks=true,citecolor=ruby,colorlinks,linkcolor=ruby,bookmarks=false]{hyperref}
\newcommand{\mm}{\textcolor[rgb]{1.0,0.2,0.4}} %% alberto

\title{Video-TransUNet: Temporally Blended Vision Transformer\\for CT VFSS Instance Segmentation}

\author[a]{Chengxi Zeng}
\author[b]{Xinyu Yang} 
\author[b]{Majid Mirmehdi}
\author[a]{Alberto M Gambaruto}
\author[b]{Tilo Burghardt}

\affil[a]{Dept of Mechanical Engineering}
\affil[b]{Dept of Computer Science}
\affil[ ]{University of Bristol, UK\vspace{-12pt}}

\authorinfo{Further author information: (Send correspondence to Chengxi Zeng)\\ E-mail: cz15306@bristol.ac.uk\\
The code is available at: \href{url}{https://github.com/SimonZeng7108/Video-TransUNet} }

\begin{document} 
\maketitle

\begin{abstract}
We propose Video-TransUNet, a deep architecture for instance segmentation in medical CT videos constructed by integrating temporal feature blending into the TransUNet deep learning framework. In particular, our approach amalgamates strong frame representation via a ResNet CNN backbone, multi-frame feature blending via a Temporal Context Module (TCM), non-local attention via a Vision Transformer, and reconstructive capabilities for multiple targets via a UNet-based convolutional-deconvolutional architecture with multiple heads. We show that this new network design can significantly outperform other state-of-the-art systems when tested on the segmentation of bolus and pharynx/larynx in Videofluoroscopic Swallowing Study (VFSS) CT sequences. On our  VFSS2022 dataset it achieves a dice coefficient of $0.8796$ and an average surface distance of $1.0379$ pixels. Note that tracking the pharyngeal bolus accurately is a particularly important application in clinical practice since it constitutes the primary method for diagnostics of swallowing impairment. Our findings suggest that the proposed model can indeed enhance the TransUNet architecture via exploiting temporal information and improving segmentation performance by a significant margin. We publish key source code, network weights, and ground truth annotations for simplified performance reproduction.
%Challenges occur for therapists with different levels of experience and distinct evaluation guidelines to agree on a consistent outcome. Additionally, due to the rapid movement of bolus and hardly perceptible laryngeal structures, the ambiguity of pharyngeal distinguishment remains for the individual image frames. With limited multi-rater fused annotated data, The model can differentiate bolus and pharynx/larynx, fully exploiting past and future relations, achieving a dice coefficient of $0.8796\%$ and an average surface distance of $1.0379$ pixels. The findings suggest that the proposed model, incorporating temporal characteristics, improves segmentation accuracy by a significant margin compared with the other models using spatial features only. 
\end{abstract}

\keywords{TransUNet, Videofluoroscopy, Dysphagia, Temporal Tracking, Vision Transformer, Deep Learning}

\section{Motivation and Introduction} \label{sec:intro}  % \label{} allows reference to this section

\textbf{Clinical Motivation}. Dysphagia is defined as a deglutition-related malfunction during human processing of solid or liquid substances from mouth to pharynx and into the esophagus. Patients who have dysphagia often suffer from swallowing discomfort and subsequently difficulty in forming or delivering food to the digestive  system\cite{Sura2012DysphagiaIT}. The condition is often accompanied with complications including severe coughing and aspiration pneumonia, which can cause underlying consequences such as suffocation, bronchitis, and chronic lung disease \cite{Smithard1996ComplicationsAO,doggett2001prevention}. Many clinical conditions lead patients to have dysphagia. It can be explicit factors such as Globus Pharyngeus~(lump in throat), cerebrovascular accident~(stroke), throat surgery, traumatic brain injury, Parkinson’s disease and Parkinson-like syndromes or nerve system diseases related to Amyotrophic Lateral Sclerosis and Multiple Sclerosis\cite{Rugiu2007RoleOV} . The high prevalence of dysphagia amongst old people has been recognised in many regions. Approximately 17\% of adults over 65 years old in Europe and 25\% - 50\% of people over 60~(that is 1 in 25 among adult citizens) in the United States suffer from different levels of dysphagia \cite{Mann2000SwallowingDF, Rofes2011DiagnosisAM} with particular prevalence in nursing homes or health care facilities~\cite{Clav2004ApproachingOD, Aslam2013DysphagiaIT}. Noting that a delayed therapy and inaccurate identification may result in fatal incidents (silent aspiration). Physical and medical imaging examinations are critical for diagnosis and are ideally maintained along rehabilitation therapies~\cite{Hinchey2005FormalDS}.

\begin{figure} [ht]
\begin{center}
\begin{tabular}{c} \vspace{-15pt}
\includegraphics[height=7cm]{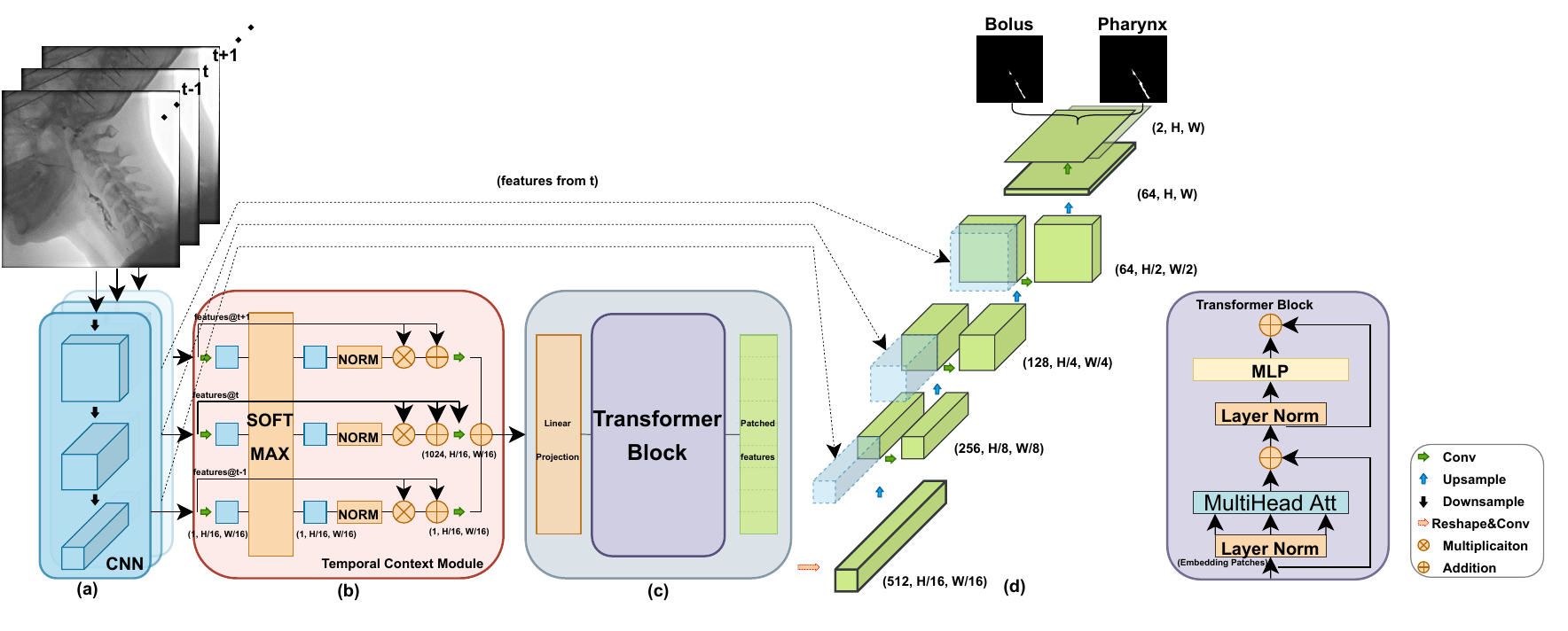}
\end{tabular}
\end{center}
\caption[example] 
{\label{fig:tcm_transunet} 
\textbf{Video-TransUNet Architecture Overview.} (a) Multi-frame ResNet-50-based feature extractor; (b) Temporal Context Module for temporal feature blending across frames; (c) Vision Transformer Block for non-local attention-based learning of multi-frame encoded input; (d) Cascaded expansive decoder with skip connections as used in original UNet architectures, however, here with multiple prediction heads co-learning the two  instances of clinical interest.}
\end{figure} 

 \noindent\textbf{Computer Vision Challenge}. Videofluoroscopic Swallowing Studies (VFSS) are considered the gold standard for the assessment of swallowing disorders. This X-ray imaging technique is based on modified barium swallows: a food bolus ranging across different thicknesses and textures consisting of barium are usually given and the swallowing process is analysed via computer tomography (CT) videos~\cite{MartinHarris2000ClinicalUO,spieker2000evaluating}. The resulting image sequences allow pathologists to inspect any abnormality in oral, pharyngeal, or esophageal anatomy and enable clinicians to investigate the pharyngeal residue and potential laryngeal~(airway) invasions. During the swallowing process, characteristics of anatomical structures and bolus motions are key aspects for doctors to specify the aetiology and initiate treatment plans. However, there exist challenges when analysing the videofluoroscopic videos: i) image quality is often compromised due to radiation noise and the overlapping nature of black and white X-ray scans~\cite{Zheng2004AutomatedSO}; ii) human judgment is often inconsistent on distinguishing pharyngeal and laryngeal closures~\cite{McCullough2000InterAI}; and iii) frame-by-frame human expert annotation is tedious and costly~\cite{Bini2018QuantitativeAO}. 
 
 \noindent\textbf{Recent Approaches}. Computer vision research has largely focused on hyoid bone tracking and analysing its position and displacement closely related to the functionality of the epiglottis~\cite{Paik2008MovementOT, Steele2011TheRB}. Yet, these bone dynamics vary significantly between individuals complicating the standardisation of any quantification of displacement. Most existing techniques employ traditional image processing techniques to implement hyoid bone tracking~\cite{Zheng2004AutomatedSO,Kellen2009ComputerAssistedAO} whilst a few works exist that employ deep neural networks including CNN and RNN approaches to track the hyoid bone~\cite{Zhang2018AutomaticHB, Mao2019NeckSH}, rare spatio-temporal network classifiers that detect abnormal swallows~have been tested, but usually operate without any explicit segmentation of the bolus~\cite{9098510}. Only a few recent papers have shifted work towards accurate bolus segmentation during the swallowing process ~\cite{Caliskan2020AutomatedBD, Zhang2021DeepLearningBased, bandini2021automated}. 

 \noindent\textbf{Paper Concept}. In this paper, we introduce a new deep end-to-end neural network architecture termed Video-TransUNet~(see Fig.~\ref{fig:tcm_transunet}) which, as we will show, can improve segmentations of bolus and pharynx/larynx in CT VFSS sequences beyond previous levels of accuracy. Our approach improves on existing state-of-the-art approaches by combining elements for strong frame content representation, non-local attention, and UNet-like convolutional-deconvolutional output generation with multiple heads, with multi-frame temporal feature blending. Intuitively, all information aspects related to the above components are utilised by experts during their analysis of VFSS imagery and, thus, one may hypothesise that their interplay can indeed enhance methods that do not utilise all these information, in particular single frame approaches.

  \noindent\textbf{Main Contributions}. i) \textit{Reliably Annotated Dataset}. We recruited two speech and language therapists and three trained annotators, to provide pixel-accurate segmentation of both Bolus and Pharynx for the VFSS2022 dataset, the annotated label is then fused utilising the STAPLE strategy~\cite{Warfield2004SimultaneousTA} for highest reliability. ii) \textit{Temporally Enhanced Transformer UNet}. For the first time we extend a UNet-like architecture (i.e. TransUNet) via temporal feature blending that exploits information from across neighbouring frames via attention for more accurate segmentation. iii) \textit{Detailed Experimental Evaluation}. We benchmark our system in extensively against current state-of-the-art frameworks and across various statistics to confirm widely improved performance. We also perform detailed ablations and show that our approach outperforms a single frame version of the system.

\section{Related Work}
\label{sec:Related Work}
  \noindent\textbf{Early and Related Systems}. Before the deep learning era researchers used Hough transforms~\cite{Zheng2004AutomatedSO}, Sobel Edge detection~\cite{Kellen2009ComputerAssistedAO} and Haar classifiers~\cite{Noorwali2013SemiautomaticTO} to facilitate segmentation in the VFSS domain. All methods were semi-automatic utilising manual pre-processing or hand-crafted feature extraction. Deep learning models have dramatically enhanced the accuracy of detection and improved the reliable applicability of automation for medical image segmentation. For the task at hand, Lee et al. published a first system to analyse the swallowing phase obtained from electrocardiograms(ECG) devices and speech signals based on three-layer ANNs \cite{Lee2009SwallowSW}, and later suggested a pipeline for extracting optical flow from RGB inputs for detecting the pharyngeal phase with a CNN-based neural network~\cite{Lee2018DetectionOT}. More recently, improvements have been made by leveraging  efficient data collection methods and the enhancement of the related CNN architecture \cite{Lee2019AutomaticDO}, empowering the automatic measurement of the pharyngeal swallowing time reflex from the VFSS videos~\cite{Lee2020MachineLA}. However, the majority of Lee et al.'s work is solely concerned with the pharyngeal phase or reflex time of patients. Zhang et al.~\cite{Zhang2018AutomaticHB} employed deep learning techniques such as a single shot multibox detectors to localise  the hyoid bone. An extra sensor on the throat is required in their work and the relationship between the hyoid bone displacement and sensor signals is studied by an RNN model~\cite{Mao2019NeckSH}. Other researchers published work for the recognition and segmentation of cervical vertebrae~\cite{Zhang2021AutomaticAO} and chewing structures~\cite{Iyerchewing} using deep neural networks, however, none of these works attempt segmenting the bolus or similarly dynamic liquids. 
 
  \noindent\textbf{Bolus Segmentation}. Related works on segmenting a food bolus in CT scans is very limited. Caliskan et al.~\cite{Caliskan2020AutomatedBD} trained a Mask-RCNN network on swallowing data to detect and segment the bolus, achieving an IoU of 0.71 segmentation accuracy on their datasets. Recently, Bandini et al.~\cite{bandini2021automated} used methods such as Grad-CAM and Active Contour algorithms to localise the bolus in a weakly-supervised manner. Zhang et al.~\cite{Zhang2021DeepLearningBased} compared four different deep neural network models to segment vallecular bolus residues and achieved 0.72 segmentation accuracy with the ensemble architecture. None of these existing bolus segmentation approaches provide either data or code base publicly, thus direct performance comparison is not straight forward. In contrast, we provide our code base and also network weights for full comparability.
 
  \noindent\textbf{Instance Tracking}. The inherent video format of VFSS data does not only have appearance features but also implicit temporal relationships. A large body of works well beyond this review addresses tracking. Focusing on a few recent ideas, T-CNN proposed the concept of integrating temporal and spatial information from tubelets of videos~\cite{Kang2018TCNNTW}. Optical flow approaches estimate the vector field of pixel motion to build temporal relationships. Optical flow-based neural networks by~\cite{Dosovitskiy2015FlowNetLO, Sun2018PWCNetCF} and FGFA\cite{Zhu2017FlowGuidedFA} process spatial-temporal information, the latter using weighted fusion of neighbouring features powered by optical flow. Optical flow and frame-by-frame tracking, however, only link appearance information between frames. For video instance segmentation we would like to combine information from entire sequences to segment a central frame.
 
   \noindent\textbf{Attention-based Feature Blending}. We will follow the approach of attention-based multi-frame feature composition, where the attention mechanism was advocated and popularised for instance in Vaswani et al.~\cite{Vaswani2017AttentionIA}. Cao et al.~\cite{Cao2019GCNetNN} introduced a first feature blending network and Yang et al.~\cite{Yang2019GreatAD} proposed an attention-based spatio-temporal blending framework for video. Originally used to detect part-occluded and camouflaged animals in jungle camera trap settings\mm{,} the framework aggregates information from an entire frame stack to infer information of a single frame. The work used a feature pyramid network as a feature extractor and then fused the features from past and future frames into a single representation relating to one target frame. Our work is heavily based on this idea as well as on taking advantage of UNet~\cite{ronneberger2015unet} segmentation strategy that combines advantages of CNNs and vision transformers~\cite{Chen2021TransUNetTM}. To the best of our knowledge, our Video-TransUNet approach is the first to combine a temporal attention mechanism for video with a transformer-enhanced UNet -- no such approach has been trialled before on medical data such as VFSS in particular. 

\section{Method}
\label{sec:Method}
 \noindent\textbf{Conceptual Overview.} Our work follows the core pipeline of single image UNets~\cite{ronneberger2015unet} yet accepts an entire frame stack of video~(see Fig.~\ref{fig:tcm_transunet}(a)), that is utilising inputs $\mathbf{x} \in \mathbb{R}^{t \times H \times W}$ with a spatial resolution of $H \times W$ and snippet length of $t$. Features are extracted from each frame of this stack via a ResNet-50 backbone CNN~\cite{He2016DeepRL} producing $t$ abstract features. Our key novel enhancement of the UNet/Trans-UNet concepts rests with blending all these extracted features from the past and future frames into a central representation turning the system into a true video-processing framework. We integrate a Temporal Context Module~(TCM) component originally used in detection pipelines~\cite{Yang2019GreatAD} into the UNet concept to achieve this~(see Fig.~\ref{fig:tcm_transunet}(b)). Following the Trans-UNet ideas~\cite{Chen2021TransUNetTM}, the resulting representation forms the input to a transformer encoder block to learn high-level representations~(see Fig.~\ref{fig:tcm_transunet}(c)). Finally, the decoder generates full-size $H \times W$ binary segmentations through cascaded up-sampling and various segmentation heads~(see Fig.~\ref{fig:tcm_transunet}(d)). The final output has the shape of $\mathbf{x} \in \mathbb{R}^{2 \times H \times W}$ since both the bolus label and pharynx label are generated.
 
 \noindent\textbf{Temporal Feature Blending.} Following~\cite{Yang2019GreatAD}, we inserted a TCM shown in Fig.\ref{fig:tcm_transunet}(b) into the Trans-UNet pipeline. It is a trainable self-attention module that compacts a feature representation based on learned attention across several frame features. In particular, the features $x_t$ extracted from previous CNN layers are linearly embedded by a function~$e(\cdot)$ and multiplied by weights. Thereafter, the temporal correlation between all $T$ frames is calculated by a global \texttt{Softmax} operation, thus the nearby features are temporally blended across and cast into a single, task-effective representation as
\begin{equation}
\mathcal{C}\left(x_{t, i} ; w\right)=\frac{\exp \left(w_{t} x_{t, i}\right)}{\sum_{n \in T} \exp \left(w_{n} x_{n, i}\right)}~,
\end{equation}
where $i$ is the temporal position index of the features, $x_t$ represents the features from an incoming frame, and $x_n$ enumerates all the features. The $w$s are the linear embedding parameters for corresponding time frames.
Normalisation of the correlation values is necessary and is done simply by: $\hat{x}_{t, i}=\frac{1}{H W} \mathcal{C}\left(x_{t, i} ; w_t\right) \sum_{j=1}^{H W} \mathcal{C}\left(x_{t, j} ; w_t\right)$. The normalised feature maps are subsequently both multiplied and added back to the original features individually with a new set of weights providing enhanced learning stability~\cite{Yang2019GreatAD}. A final step blends all weighted and stabilised frame-features into a single mixture representation with learned weights as
\begin{equation}
z_{t, i}^{T C M}=x_{t, i}+\sum_{n \in T} w_{n}^{**}\left(x_{n, i} \oplus w_{n}^{*} \sum_{j=1}^{HW} \hat{x}_{n, j} \otimes x_{n, j}\right)~,
\end{equation}\\
where $x_n$ are the features to be blended,and $w_n^{**}$ are trainable adaptive weights and the two operations, $\oplus$ and $\otimes$, are matrix addition and multiplication respectively.

 \noindent\textbf{Transformer-based Feature Encoding.} The temporally blended features are fed into a transformer encoder in order to learn task-relevant content relationships right at the bottleneck of the U-shaped architecture. As detailed in~\cite{Chen2021TransUNetTM}, for this component features are mapped to an output $z_{0}$ via learning a patch embedding to a latent N-dimensional embedding space with positional information~$E_{\text {pos}}$ added, such that 
\begin{equation}
z_{0}=\left[x_{p}^{1} E_{pat} ; x_{p}^{2} E_{pat} ; \cdots ; x_{p}^{N} E_{pat}\right]+E_{\text {pos}} ~. 
\end{equation}\\
For a whole transformer block this mapping and the block details are described with all specifics for instance in~\cite{Dosovitskiy2021AnII}. Fig.\ref{fig:tcm_transunet}(right) illustrates the block. The key components are linked to normalisation of L layers via Multi-head Self-Attention~(MSA) followed by another normalisation and a final Multi-Layer Perceptron~(MLP), i.e.
\begin{equation}
\begin{aligned}
&z_{\ell}^{*}=\operatorname{MSA}\left(\mathrm{LN}\left(z_{\ell-1}\right)\right)+z_{\ell-1} ~, \\
&z_{\ell}=\operatorname{MLP}\left(\operatorname{LN}\left(z_{\ell}^{\prime}\right)\right)+z_{\ell}^{*} ~,
\end{aligned}
\end{equation}
where $\mathrm{LN}({\cdot})$ represents layer normalisation. Following the UNet~\cite{ronneberger2015unet} principle, the resulting feature is then ready to be expanded in the decoder utilising further skip connections.\\
\textbf{Decoder.} With three cascaded \texttt{up-sample} and \texttt{convolution} operations taking into account skip connection input~\cite{ronneberger2015unet}, we finally generate a full size output of $H \times W$ given as $\mathbf{z}_{L} \in \mathbb{R}^{\frac{H W}{P^{2}} \times D}$. As mentioned, each of the up-sampling stages is skip-connected with features from CNN encoding the frame segmentation is predicted for. The main benefit for passing the perceived semantic and low-level features from shallow layers to the layer stage is that finer details can be captured in complex background as shown in the ablation study of~\cite{Chen2021TransUNetTM}. Multiple heads are used to produce outputs for different target instances. 

\noindent We will now describe experiments conducted with this architecture to demonstrate its efficacy, evaluate performance in comparison to other state-of-the art segmentation approaches and ablate key design elements.

\section{Experiments and Results}
\subsection{Dataset}
\label{sec:Datasets}
\begin{figure}[t]
\begin{center}
\begin{tabular}{c} 
\includegraphics[height=3.5cm]{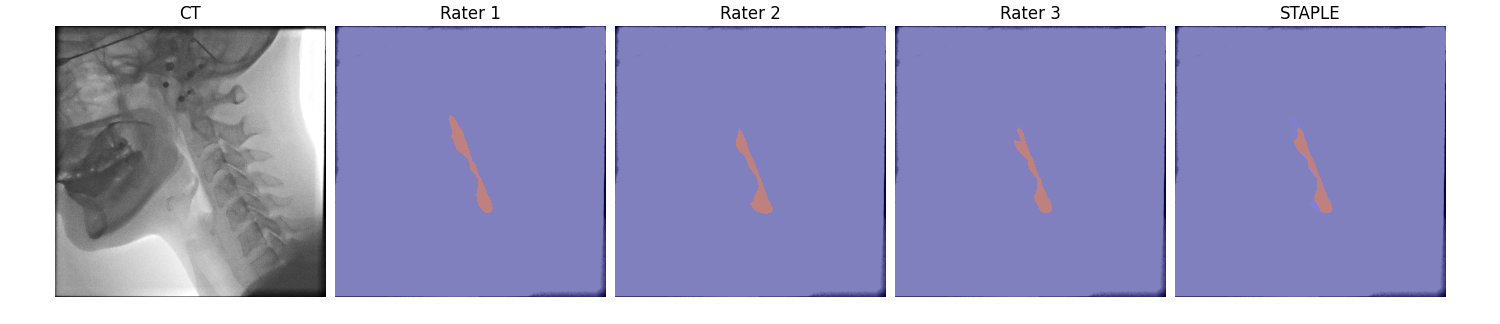}
\end{tabular}
\end{center}
\caption[example] 
{\label{fig:staple} 
\textbf{VFSS DATA AND GROUND TRUTH:} (left) Sample VFSS frame, one amongst 440 such $512\times 512$ greyscale images generated via CT. (middle) Three annotations for segmentation provided by experts. (right) Data fusion via the STAPLE~\cite{Warfield2004SimultaneousTA} algorithm was used to generate a single, more reliable ground truth annotation. This algorithm is applied to bolus annotations and phayrnx annotations separately.}
\end{figure}

The imagery underpinning our study was collected externally at the Southmead Hospital at Bristol, UK. 
The utilisation of these anonymised data was reviewed and approved at source, but also by our internal institutional Ethics Committee (REF: 63982). Patients conducted standard Barium swallow tests under clinicians' supervision.  A total of 17 swallows were performed and measured via VFSS CT imagery leading to overall 440 utilised frames at a resolution of $512\times 512$ pixels comprising the VFSS2022 dataset. Fig.~\ref{fig:staple} depicts an example frame together with the three expert annotations and our STAPLE-generated most reliable ground truth. The data were annotated by three trained experts and certified by two speech pathologists. Each rater produced pixel-grade bolus and pharynx annotations for each frame. The three sets of annotations are fused by the STAPLE algorithm designed to produce more reliable prototype annotations given slightly varying annotator input~\cite{Warfield2004SimultaneousTA}. The resulting dataset with annotations was split into three portions, training set(70\%), validation set(15\%), and test set(15\%).\\

\subsection{Implementation Details}
\label{sec:Implementation Details}
 Bolus and pharynx annotations are concatenated together as a two-layer tensor ground truth to be evaluated against the two-layer output from the proposed end-to-end system. For images that did not contain any bolus, a full-size zeroed image represents this situation. For experiments utilising the TCM component, we choose snippet lengths of $t = 3, 5, 7, 9, 11, 13$ for training and $t = 3, 5, 7, 9, 11, 13, 15, 17$ for testing. All experiments utilised data augmentations for training covering limited random rotation and flipping. The patch size for linear embeddings is set to $16$. The ResNet-50 backbone CNN and ViT are initialised with the pre-trained weights from ImageNet~\cite{Dosovitskiy2021AnII}. The whole model is trained with the Adam optimiser with an initial learning rate of $5e^{-4}$. The learning rate is scheduled to half after 20 epochs one performance saturation is reached (i.e. the moment validation has no loss decrease). A batch size of 2 was chosen and all experiments were run on an NVIDIA Tesla P100 16 GB GPU. All experiments were conducted under the environment of Python 3.8.5 and Pytorch 1.9. The supplied code shows how input tensors are adjusted in detail and how the training loss function averaged across both outputs balances Binary Cross Entropy loss, Dice loss, and Hausdorff Distance Loss via a mixture. The evaluation metrics we used for testing are the Dice Coefficient~(DSC), the 95th percentile of the Hausdorff Distance~(HD95), the Average Surface Distance~(ASD), and the more general measures of Sensitivity and Specificity. \\

\begin{table}[ht]
\caption{\textbf{Quantitative Results} Dice Coefficient~(DSC), 95th percentile of the Hausdorff Distance~(HD95), Average Surface Distance~(ASD), as well as Sensitivity and Specificity are provided for the VFSS2022 test set for our proposed approach and across other six state-of-the-art segmentation techniques. Note the dominance of our TCM utilisation approach (6) across most measures and the significant performance leap regarding HD95 and ASD that transformer-included pipelines (5) and (6) can provide.} 
\label{tab:comparison method}
\begin{center}       
\begin{tabular}{l|l|l|l|l|l|l} 
\toprule
\rule[-1ex]{0pt}{3.5ex}  \textbf{Model} & \textbf{Year} & \textbf{DSC} & \textbf{HD95} & \textbf{ASD} & \textbf{Sensitivity} & \textbf{Specificity} \\
\midrule \midrule
\rule[-1ex]{0pt}{3.5ex}  (1) UNet\cite{ronneberger2015unet} & $2015$ & $0.8422$ & $14.7530$ & $2.1675$ & $0.8289$ & $0.9988$  \\

\rule[-1ex]{0pt}{3.5ex}  (2) NestedUNet\cite{Zhou2018UNetAN} & $2018$ & $0.8335$ & $13.7601$ & $2.2275$ & $0.8305$ & $0.9987$  \\

\rule[-1ex]{0pt}{3.5ex}  (3) ResUNet\cite{Zhang2018RoadEB} & $2018$ & $0.8465$ & $11.9820$ & $2.0487$ & $0.8183$ & $\mathbf{0.9991}$ \\

\rule[-1ex]{0pt}{3.5ex}  (4) AttUNet\cite{Oktay2018AttentionUL} & $2018$ & $0.8501$ & $12.9356$ & $2.1832$ & $0.8328$ & $0.9988$   \\

\rule[-1ex]{0pt}{3.5ex} (5)  TransUNet\cite{Chen2021TransUNetTM} & $2021$ & $0.8586$ & $7.4510$ & $1.1050$ & $0.8486$ & $0 . 9 9 8 9$  \\

\rule[-1ex]{0pt}{3.5ex}  (6) Video-TransUNet~(Ours) & 2022 & $\mathbf{0 . 8 7 9 6}$ & $\mathbf{6 . 9 1 5 5}$ & $\mathbf{1 . 0 3 7 9}$ & $\mathbf{0 . 8851}$ & $0.9986$  \\
\midrule
\end{tabular}
\end{center}
\end{table}

\subsection{Results}
\begin{figure*}[ht]
\centering
\includegraphics[width=6.8 in]{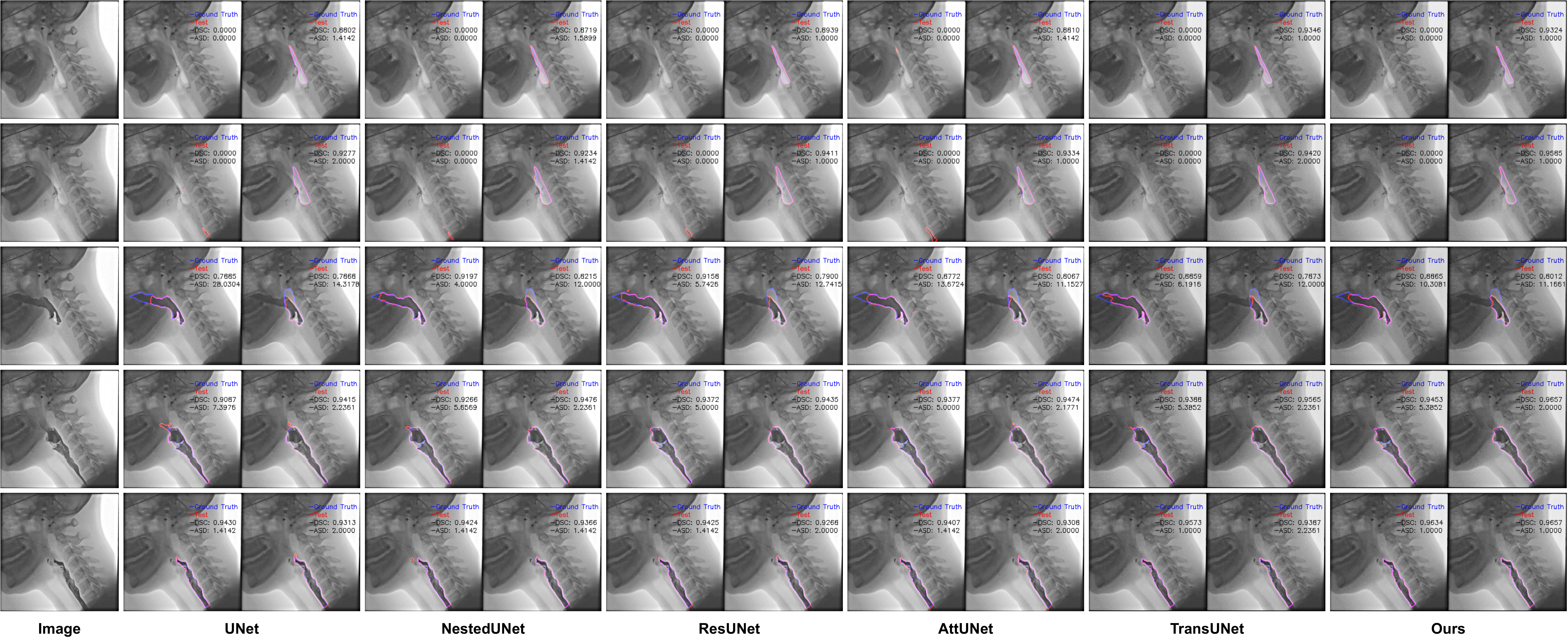}
\caption{\textbf{Qualitative Results.} Visualisation of segmentation outputs for five selected frames (one frame per row) from a VFSS test video sequence. (left column) original frame (other columns) segmentation results of all tested methods with bolus and pharynx results side by side. The ground truth of instance boundaries is shown in blue and the automatically produced segmentation in red. Purple pixels depict coincidence of ground truth and result pixels. (best viewed zoomed-in)}
\label{fig: qualitative}
\end{figure*}

Tab.~\ref{tab:comparison method} above summarises quantitative results comparing our proposed architecture via a multitude of key statistics in row (6) to other recent state-of-the-art segmentation approaches for medical images including (1) Unet\cite{ronneberger2015unet}; (2)N estedUnet~\cite{Zhou2018UNetAN}; (3) ResUNet~\cite{Zhang2018RoadEB}; (4) AttUNet~\cite{Oktay2018AttentionUL} and (5) TransUNet~\cite{Chen2021TransUNetTM}. For fairest comparison, we use ResNet-50 as the backbone for ResUNet as used in our work showing the effectiveness of adding TCM and transformer blocks in particular. It can be seen in the table that the proposed approach widely dominates benchmarks for all compound measures (that is DSC, HD95, ASD). It also shows very strong sensitivity close to $4\%$ ahead of other methods. Fig.~\ref{fig: qualitative} depicts qualitative test results on a few sample frames across the six tested methods with bolus and pharynx segmentation outputs shown side by side.  

\subsection{Result and Ablation Discussion}
\begin{figure*}[t]
\centering
\includegraphics[width=4.5 in]{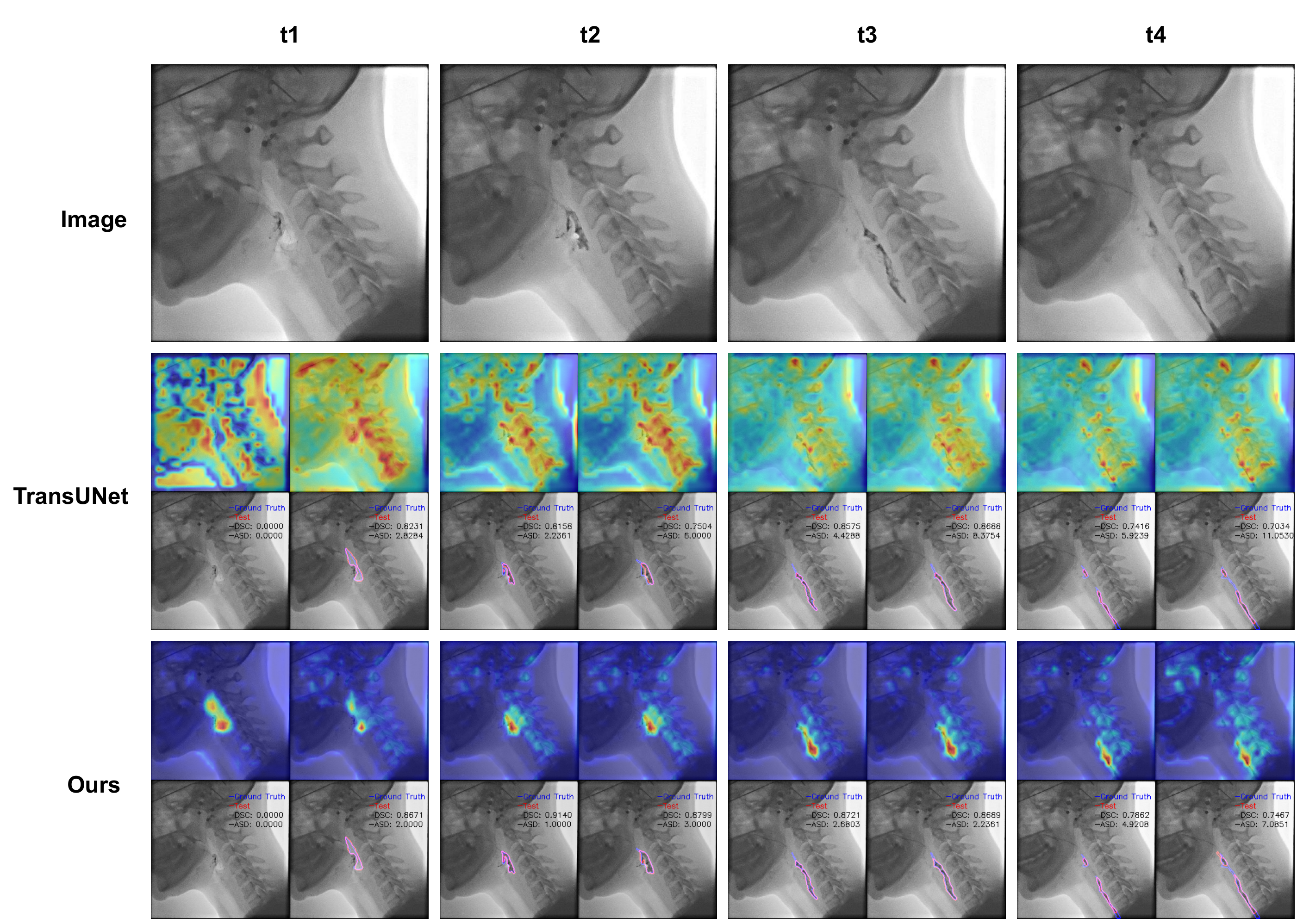}
\caption{\textbf{Attention Visualisation.} Based on four sample frames (top) we show for TransUNet and our model boundary segmentations (lower rows) and GradCam output (upper rows) highlighting where models are paying attention to. Results for the bolus and pharynx are next to each other left and right, respectively, for every sample image. Note the much more target-focused results of our model.}
\label{figure:grad_cam}
\end{figure*}

The dice coefficient (DSC) reveals the general overlap between ground truth and output sets providing the most direct reflection of model performance in structural segment reproduction. Our method provides a significant improvement of 2.1\% DSC compared to the next competitive approach (TransUNet) on the dataset. This finding emphasises the importance of considering temporal information during segmentation since the TCM insertion and resulting multi-frame feature fusion is the key and only structural difference between these two network designs.  Fig.~\ref{figure:grad_cam} shows the GradCam output of four consecutive frames just one layer before the Transformer block in the CNN encoder comparing TransUNet and the exact same feature location in our model. The resulting attention maps show a conspicuous difference between the two models: they confirm the effectiveness of our TCM component insertion to capture task-relevant information via attention and focus network attention well to the relevant spatial regions in the frame of interest. The other two distance metrics (HD95, ASD) confirm model superiority across other measures with good stability. AttUNet can be considered a basic spatial feature blender with strength in computing long-range high-level semantics for segmentation maps. Consequently, it achieved good results of $85.01\%$ DSC only beaten by TransUNet and our model. Noticeably, both our model and TransUNet yield some significant advancement on the distance metrics credit to the transformer block added to the UNet pipelines. Note that since TransUNet can be seen as identical to our model once multi-frame information and its blending via the TCM are removed, the performance improvements from row (5) to row (6) in Tab.~\ref{tab:comparison method} provide a whole-component ablation for the use of the TCM subnet. We will now look at qualitative aspects and parameter studies and ablations further to justify our approach taken. 

\subsection{Qualitative Result Discussion} For Fig.~\ref{fig: qualitative} we selected 2 scans without bolus~(rows 1 and 2) and 3 frames from a sequence (rows 3, 4 and 5) covering the core swallowing process in order to outline some typical features of each of the methods tested. Generally, all models can differentiate bolus and pharynx well despite spatial overlap between the two.
In some cases, e.g. row 2, models without a transformer block are not able to determine the presence of a bolus accurately and sometimes produce false positive detections. TransUNet and our model omit such scenarios yet still capture fine details for the pharynx. Beyond false positives, UNet and NestedUnet do not generate smooth segmentation maps as they lack more powerful encoders such as ResNet-50. However, as shown in rows 3 and 4, ResUNet and AttUNet also produce some noise compared to the ground truth resulting in lower sensitivity~(see Tab.~\ref{tab:comparison method}). In rows 3 to 5 limited under-segmentation can be observed for TransUNet compared to our model especially for pharynx segmentation highlighting a qualitative example where attention-learned temporal information is beneficial. 

\subsection{Extent of Temporal Blending}
\textbf{Snippet Length.} We carried out a further evaluation to rigorously investigate the impact of snippet length, that is the number of frames blended by the TCM component, on performance as shown in Tab.~\ref{tab:tcm train}. Note that longer snippet length in the training stage will result in a larger input size and impact the model training speed and memory footprint. It can be observed that 5 frames including the central frame where segmentation is carried out are best for training. This rather short window is intuitively understandable since the rapid swallowing process normally takes not many more than 5 frames thus surrounding frames provide little extra information regarding exact bolus and pharynx locations.
\begin{table}[t]
\caption{\textbf{Snippet Length.} DSC performance of the test section of VFSS2022 when altering the number of frames used in the TCM component.} 
\label{tab:tcm train}
\begin{center}       
\begin{tabular}{|l|l|l|l|l|l|l|} 
\hline
\rule[-1ex]{0pt}{3.5ex}  Snippet Length & 3 frames & 5 frames & 7 frames & 9 frames & 11 frames & 13 frames \\
\hline
\rule[-1ex]{0pt}{3.5ex}  DSC & $0.8570$ & $\mathbf{0.8796}$ & $0.8682$ & $0.8541$ & $0.8508$ & $0.8676$ \\
\hline
\end{tabular}
\end{center}
\end{table}

\section{Conclusion}
In this paper we introduced Video-TransUNet and applied it to instance segmentation in medical CT VFSS videos leading to significant performance gains on the VFSS2022 dataset against leading methods. We constructed Video-TransUNet by integrating temporal feature blending into a TransUNet end-to-end deep learning framework. The resulting architecture combines a strong encoder via a ResNet CNN backbone, multi-frame feature blending via a Temporal Context Module, non-local attention-based learning via a Vision Transformer, and reconstructive capabilities for multiple targets via a UNet-based deconvolutional expansion network with multiple heads. We show that this new design can significantly outperform other state-of-the-art systems when tested on the segmentation of bolus and pharynx/larynx on the VFSS2022 dataset. We ablated the effect of the multi-frame TCM approach (Tab.1 rows (5) vs (6)) and showed that the suggested feature blending indeed benefits performance. We provided detailed quantitative and qualitative insights and publish source code and network weights with this paper. We hope that our contribution can help the application field to further research the domain and utilise multi-frame information blending for single frame segmentation in deep medical image analysis more widely. 

\acknowledgments    % equivalent to \section*{ACKNOWLEDGMENTS}
The institutional Ethics Committee approved data usage and research application (REF: 11277). Regarding the dataset VSFF2022 used, the authors would like to thank Prof. David G Smithard, the speech and language therapists, Russel Walker from North Bristol NHS Trust and Aoife Stone-Ghariani from Woolwich Queen Elizabeth Hospital, for their annotation supervision and data annotation, and also Yuri Lewyckyj and Victor Perez for their work on annotations.\\
Project: Chin Tuck Against Resistance with Feedback: Swallowing Rehabilitation in Frail Older People - A feasibility study.\\
Funder: NIHR, Research for Patient Benefit.\\
Ref: PB-PG-1217-20005.

% References
\footnotesize
\bibliography{report} % bibliography data in report.bib
\bibliographystyle{unsrt} % makes bibtex use spiebib.bst

\end{document}